\documentclass[aoas,nameyear,dvips]{arximspdf}
\usepackage{multirow}
\usepackage{graphics}

\doi{10.1214/09-AOAS257}
\volume{3}
\issue{4}
\pubyear{2009}
\firstpage{1758}
\lastpage{1775}

\makeatletter
\newtheorem{theorem}{Theorem}
\makeatother

\begin{document}
\begin{frontmatter}

\title{Detecting and handling outlying trajectories in irregularly sampled
functional datasets}
\runtitle{Outlying curves in functional data}

\begin{aug}
\author[A]{\fnms{Daniel} \snm{Gervini}\thanksref{t1}\ead[label=e1]{gervini@uwm.edu}\corref{}}
\runauthor{D. Gervini}
\thankstext{t1}{Supported by NSF Grant DMS-06-04396.}
\affiliation{University of Wisconsin--Milwaukee}
\address[A]{Department of Mathematical Sciences\\
University of Wisconsin--Milwaukee\\
P.~O. Box 413\\
Milwaukee, Wisconsin 53201\\
 USA\\
\printead{e1}} 
\end{aug}

\received{\smonth{1} \syear{2009}}
\revised{\smonth{5} \syear{2009}}

%
\begin{abstract}
Outlying curves often occur in functional or longitudinal datasets, and can
be very influential on parameter estimators and very hard to detect
visually. In this article we introduce estimators of the mean and the
principal components that are resistant to, and then can be used for
detection of, outlying sample trajectories. The~estimators are based on
reduced-rank \textit{t}-models and are specifically aimed at sparse and
irregularly sampled functional data. The~outlier-resistance properties of
the estimators and their relative efficiency for noncontaminated data are
studied theoretically and by simulation. Applications to the analysis of
Internet traffic data and glycated hemoglobin levels in diabetic children
are presented.
\end{abstract}

%

\end{frontmatter}
%

\section{Introduction}\label{sec1}

In many statistical problems the collected data consists of samples of
stochastic processes rather than scalars or vectors. Typical examples
include human growth curves and circadian rhythms in medicine,
time-dependent gene expression profiles in genomics, and spectral
curves in
chemometrics. Other examples and an overview of the related statistical
methodology can be found in Ramsay and Silverman (\citeyear{RaSi2005}).

\begin{figure}

\includegraphics{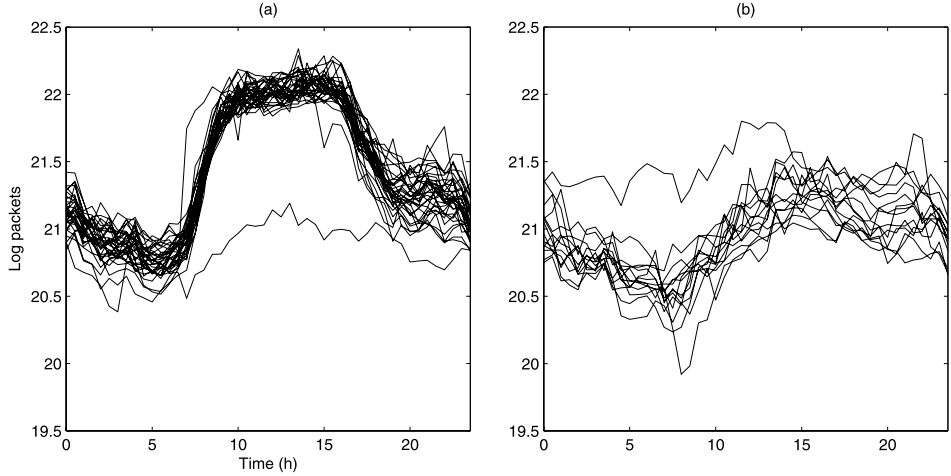}

\caption{Internet traffic data.
Trajectories for (\textup{a}) 35 week days and (\textup{b}) 14 weekend days.}\label{fig:traffic_data}\vspace*{-8pt}
\end{figure}

As with univariate or multivariate samples, the presence of atypical
observations in functional samples tends to complicate the statistical
analysis. By atypical observations we mean atypical curves, not just
isolated points. To illustrate the problem, consider the following two
examples. The~first one is a problem on Internet traffic analysis. The~data,
previously analyzed by Zhang et al.~(\citeyear{ZhMaShZh2007}), was collected at the main
Internet link of the University of North Carolina campus network during
seven consecutive weeks. The~traffic is measured in packet counts, every
half an hour; the logarithm of the data for the 35 week days is shown in
Figure~\ref{fig:traffic_data}(a). Most trajectories, while noisy, show
a clear
daily pattern: the traffic rises sharply between 7 and 9 a.m., remains at
approximately the same level between 9~a.m. and 4 p.m., and goes down again
between 4 and 7 p.m. However, there is a clearly atypical curve, a day with
unusually low traffic, and another one less conspicuous but still atypical,
corresponding to a day when the traffic peaked earlier than usual in the
morning. The~problems created by these atypical curves, and how to deal with
them, will be discussed more extensively in Section~\ref{sec:Example}.

\begin{figure}[b]
\vspace*{-8pt}
\includegraphics{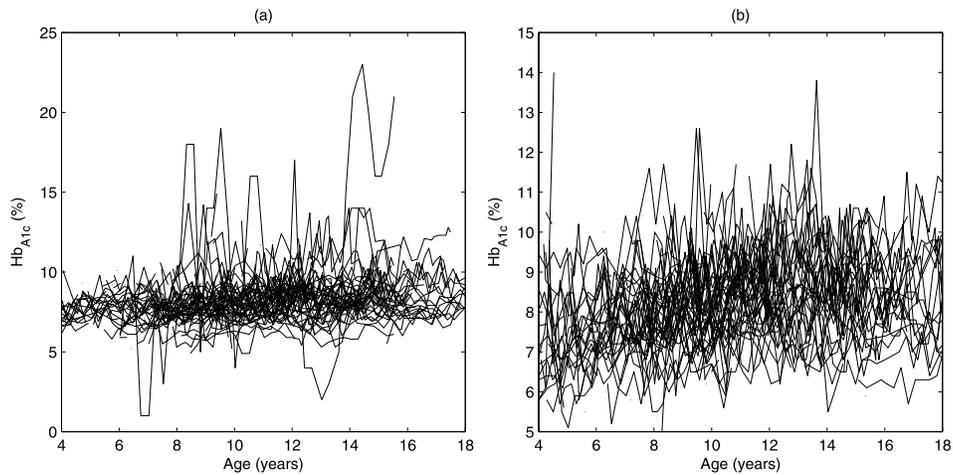}

\caption{Child diabetes data.
Trajectories of $Hb_{A1c}$ levels for (\textup{a}) 73 females and (\textup{b}) 66
males.}\label{fig:trajectories}
\end{figure}

The~second example is more complicated. Figure \ref{fig:trajectories} shows
trajectories of glycated hemoglobin levels for diabetic children who
underwent treatment at the Children's Hospital of the University of Zurich.
The~level of glycated hemoglobin (abbreviated $Hb_{A1c}$) is used to assess
the effectiveness of therapy in patients with type-I diabetes mellitus, and
to study the long-term effect of the disease on physical and intellectual
development [see, e.g.,~Schoenle et al. (\citeyear{ScScMoLa2002})]. One trajectory is
clearly out
of control in Figure~\ref{fig:trajectories}(a), but besides that, it
is hard
to discern any systematic patterns in the data. To complicate the
problem, $%
Hb_{A1c}$ levels are measured at irregular time points, with as few as 2
observations for some individuals. This makes individual smoothing of the
trajectories (which would have eased visualization) very hard or even
impossible. This example will also be discussed in more detail in
Section %
\ref{sec:Example}, but it is clear that atypical curves cannot always be
detected by visual inspection, and one must rely on methods that can handle
outlying curves automatically.

These examples also show that outliers, in the functional sense, are not
simply the result of misrecorded data or extreme noise. They correspond to
individuals that, for some reason, do not follow the pattern of the majority
of the data, and often deserve to be studied more carefully rather than
simply discarded. However, these atypical curves must be downweighted
at the
estimation step, or they may lead to erroneous conclusions for the rest of
the population.

This article is organized as follows. Section \ref{sec:FDmodels}
frames the
discussion in a more rigorous statistical setting, as an estimation problem
for stochastic processes. Section \ref{sec:t-estimators} proposes an
outlier-resistant estimation method, and Sections \ref{sec:Properties}
and %
\ref{sec:Simulations} discuss their asymptotic and finite-sample properties.
Section \ref{sec:Example} presents a more thorough analysis of the above
examples. Available as supplementary material are a Technical Report with
proofs and mathematical derivations, and Matlab programs implementing the
proposed estimators.

\section{Functional data models}\label{sec:FDmodels}

The~data in the examples above and in similar longitudinal studies can be
thought of as discrete observations of continuous-time stochastic processes
(or, more generally, of stochastic processes depending on a continuous
variable). Usually, the data is observed with random noise:
%
\begin{eqnarray}
x_{ij}=X_{i}(t_{ij})+\varepsilon_{ij},\qquad   j=1,\ldots,m_{i},  i=1,\ldots
,n, \label{eq:raw-data}
\end{eqnarray}
where $\{X_{i}(t)\}$ are i.i.d.~trajectories of the stochastic process of
interest, $\{t_{ij}\}$ are the time points where the trajectories are
measured, and $\{\varepsilon_{ij}\}$ are independent random errors. It is
known [see, e.g.,~Gohberg, Goldberg and Kaashoek (\citeyear{GoGoKa2003})] that a stochastic
process $X\in L^{2}([a,b])$ with $E(\Vert X\Vert^{2})<\infty$ admits the
expansion (known as Karhunen--Lo\`{e}ve decomposition)
%
\begin{eqnarray}
X(t)=\mu(t)+\sum_{k=1}^{\infty}y_{k}\phi_{k}(t), \label{eq:K-L-decomp}
\end{eqnarray}
where $\mu(t)=\mathrm{E}\{X(t)\}$. The~$\phi_{k}$s form a nonrandom
orthonormal basis of $L^{2}([a,b])$ and the $y_{k}$s are uncorrelated random
variables with zero mean and finite variance. If $\rho(s,t)=\mathrm
{cov}%
\{X(s),X(t)\}$, we have the representation
%
\begin{eqnarray}
\rho(s,t)=\sum_{k=1}^{\infty}\lambda_{k}\phi_{k}(s)\phi_{k}(t),
\label{eq:rho}
\end{eqnarray}
where $\lambda_{k}=\mathrm{var}(y_{k})$. If $\rho(s,t)$ is continuous,
then the $\phi_{k}$s are also continuous and the series (\ref{eq:rho})
converges uniformly and absolutely. This representation implies that $%
\lambda_{k}$ is an eigenvalue of $\rho$ with eigenfunction $\phi
_{k}$, so
the $\phi_{k}$s are called ``principal
components'' and the $y_{k}$s ``component
scores,'' in analogy with multivariate analysis.

To a large extent, the stochastic process $X(t)$ is characterized by
$\mu
(t) $ and $\rho(s,t)$. Estimating these functions is challenging when the
time grid $\{t_{ij}\}$ is irregular or sparse, because it makes individual
smoothing of the trajectories very hard or even impossible ($m_{i}$ may be
as low as 1 or 2 for some individuals). Some authors that have addressed
this problems are Staniswalis and Lee (\citeyear{StLe1998}), Yao, M\"{u}ller and Wang
(\citeyear{YaMuWa2005}), James, Hastie and Sugar (\citeyear{JaHaSu2000}), Gervini (\citeyear{Ge2006}), and Yao and Lee
(\citeyear{YaLe2006}). These estimators, however, cannot handle outlying curves like those
in the examples of the Introduction. Estimators that do handle outlying
curves were proposed by Locantore et al.~(\citeyear{LoMaSiTrZhCo1999}), Fraiman and Muniz (\citeyear{FrMu2001}),
Cuevas, Febrero and Fraiman (\citeyear{CuFeFr2007}), and Gervini (\citeyear{Ge2008}), but they can
only be
applied to individually smoothed trajectories. Estimators that are able to
handle outlying curves and can be computed on sparse and irregularly sampled
data have not yet been proposed. We present one possible approach in the
next section.\looseness=1

\section{Reduced-rank $t$-models}\label{sec:t-estimators}

The~eigenvalues $\lambda_{k}$ in (\ref{eq:rho}) typically decrease to zero
very fast, because $\sum_{k=1}^{\infty}\lambda_{k}<\infty$. Therefore,
only the leading terms in (\ref{eq:K-L-decomp}) are of practical relevance,
and we can assume
%
\begin{eqnarray}
X(t)=\mu(t)+\sum_{k=1}^{d}y_{k}\phi_{k}(t) \label{eq:model-for-X}
\end{eqnarray}
for some $d$, where $\lambda_{1}\geq\cdots\geq\lambda_{d}>0$.
Smoothness of $\mu$ and the $\phi_{k}$s can be built into the model by
assuming they are spline functions. That is, we assume $\mu(t)=\bolds{\theta}^{T}\mathbf{b}(t)$ and $\phi_{k}(t)=\bolds{\eta
}_{k}^{T}\mathbf{b}%
(t)$, where $\mathbf{b}(t)\in\mathbb{R}^{p}$ is a spline basis. The~observational model implied by (\ref{eq:raw-data}) and (\ref
{eq:model-for-X}%
) can be succinctly expressed as
%
\begin{eqnarray}
\mathbf{x}_{i}=\mathbf{B}_{i}\bolds{\theta}+\mathbf{B}_{i}\mathbf
{H\bolds\Lambda
}^{{1}/{2}}\mathbf{z}_{i}+\sigma\bolds{\varepsilon}_{i},\qquad   i=1,\ldots,n, \label{eq:raw-data-model}
\end{eqnarray}
where $\mathbf{B}_{i}=[b_{k}(t_{ij})]_{(j,k)}$, $\mathbf{H}=[\bolds{\eta}%
_{1},\ldots,\bolds{\eta}_{d}]$, $\bolds{\Lambda}=\operatorname{diag}(\lambda
_{1},\ldots,\lambda_{d})$, $\mathbf{z}_{i}$ is the vector of standardized
component scores and $\bolds{\varepsilon}_{i}$ are the standardized
measurement errors. If we assume a heavy-tailed distribution for
$(\mathbf{z}%
_{i},\bolds{\varepsilon}_{i})$, outlier-resistant estimators of $\mu
$ and
the $\phi_{k}$s are obtained automatically. The~reason is that, informally
speaking, heavy-tailed models ``expect'' extreme observations, which are then downweighted by the maximum likelihood
estimation process.

Specifically, we assume that $(\mathbf{z}_{i},\bolds{\varepsilon}_{i})$
has a joint multivariate $t$ distribution with $\nu$ degrees of freedom,
location parameter $\mathbf{0}$ and scatter matrix $\mathbf{I}_{d+m_{i}}$,
which we denote by $t_{\nu}(\mathbf{0,I}_{d+m_{i}})$. Then $\mathbf
{x}%
_{i}\sim t_{\nu}(\mathbf{B}_{i}\bolds{\theta,\Sigma}_{i})$, with $
\bolds{\Sigma}_{i}=\mathbf{B}_{i}\mathbf{H\bolds\Lambda H}^{T}\mathbf{B}
_{i}^{T}+\sigma^{2}\mathbf{I}_{m_{i}}$. The~maximum likelihood estimating
equations for this model, which are derived in the Technical Report,
are the
following:
%
\begin{eqnarray}\label{eq:estim_eq_1}
\sum_{i=1}^{n}\biggl(\frac{\nu+m_{i}}{\nu+s_{i}}\biggr)\mathbf
{B}_{i}^{T}%
\bolds{\Sigma}_{i}^{-1}\mathbf{(x}_{i}\mathbf{-B}_{i}\bolds{\theta)}&=&%
\mathbf{0},\\
\label{eq:estim_eq_2}(\mathbf{I}_{d}\mathbf{-JHH}^{T})\mathbf{S}_{n}\bolds{\eta
}_{k}&=&0,\qquad  k=1,\ldots,d,\\
\label{eq:estim_eq_3}\bolds{\eta}_{k}^{T}\mathbf{S}_{n}\bolds{\eta}_{k}&=&0,\qquad
k=1,\ldots,d,
\end{eqnarray}\vspace*{-16pt}
\begin{eqnarray}
-\frac{1}{2}\sum_{i=1}^{n}\operatorname{tr}(\bolds{\Sigma
}_{i}^{-1})+\frac{1}{2}%
\sum_{i=1}^{n}\biggl(\frac{\nu+m_{i}}{\nu+s_{i}}\biggr)(\mathbf
{x}_{i}%
\mathbf{-B}_{i}\bolds{\theta})^{T}\bolds{\Sigma
}_{i}^{-2}(\mathbf{x}_{i}%
\mathbf{-B}_{i}\bolds{\theta})=0, \label{eq:estim_eq_4}
\end{eqnarray}
where
\[
\mathbf{S}_{n}=\sum_{i=1}^{n}\biggl\{-\mathbf{B}_{i}^{T}\bolds{\Sigma}%
_{i}^{-1}\mathbf{B}_{i}+\biggl(\frac{\nu+m_{i}}{\nu+s_{i}}
\biggr)\mathbf{B}%
_{i}^{T}\bolds{\Sigma}_{i}^{-1}\mathbf{(x}_{i}\mathbf
{-B}_{i}\bolds{%
\theta)(x}_{i}\mathbf{-B}_{i}\bolds{\theta)}^{T}\bolds{\Sigma
}_{i}^{-1}%
\mathbf{B}_{i}\biggr\},
\]
$s_{i}=\mathbf{(x}_{i}\mathbf{-B}_{i}\bolds{\theta)}^{T}\bolds{\Sigma}%
_{i}^{-1}\mathbf{(x}_{i}\mathbf{-B}_{i}\bolds{\theta)}$ and
$\mathbf{J}%
=[\int b_{i}(t)b_{j}(t)\,dt]_{(i,j)}$. The~best linear predictor of
$\mathbf{z}%
_{i}$ is $\mathrm{E}(\mathbf{z}_{i}\mathbf{|x}_{i})=\bolds{\Lambda
}^{{%
1}/{2}}\mathbf{H}^{T}\mathbf{B}_{i}^{T}\bolds{\Sigma
}_{i}^{-1}\mathbf{(x}%
_{i}\mathbf{-B}_{i}\bolds{\theta)}$, and $\hat{\mathbf{z}}_{i}$ is
obtained by replacing the model parameters with their estimators.

What makes these estimators robust are the weights $(\nu+m_{i})/(\nu
+s_{i}) $ that appear in equations (\ref{eq:estim_eq_1})--(\ref%
{eq:estim_eq_4}). Since $s_{i}$ is the squared Mahalanobis distance
between $%
\mathbf{x}_{i}$ and the expected trajectory $\mathbf{B}_{i}\bolds{\theta}$%
, atypical trajectories are downweighted and do not seriously affect the
estimators. Downweighting is strongest for the Cauchy model ($\nu=1$) and
becomes less pronounced as $\nu$ increases. When $\nu\rightarrow
\infty$,
$(\nu+m_{i})/(\nu+s_{i})\rightarrow1$ and one obtains the estimating
equations for the Normal reduced-rank model [James, Hastie and Sugar (\citeyear{JaHaSu2000})],
which gives equal weight to all sample curves and then lacks robustness.

These estimators can be easily computed via the EM algorithm, which is
derived in detail in the Technical Report. The~recursive steps are the
following: given current estimates $\hat{\bolds{\theta}}^{\mathrm
{old}}$, $\hat{\bolds{\Xi}}^{\mathrm{old}}$ (where $\bolds{\Xi}=\mathbf{H}\bolds\Lambda^{1/2}$)
and $(\hat{\sigma}^{2})^{\mathrm{old}}$, the updates are
\begin{eqnarray*}
\hat{\bolds{\theta}}^{\mathrm{new}}&=&\Biggl\{\sum_{i=1}^{n}
\biggl(\frac{\nu
+m_{i}}{\nu+\hat{s}_{i}^{\mathrm{old}}}\biggr)\mathbf
{B}_{i}^{T}\mathbf{B}%
_{i}\Biggr\}^{-1}\sum_{i=1}^{n}\biggl(\frac{\nu+m_{i}}{\nu+\hat
{s}_{i}^{%
\mathrm{old}}}\biggr)\mathbf{B}_{i}^{T}(\mathbf{x}_{i}\mathbf
{-B}_{i}\mathbf{%
\hat{\Xi}}^{\mathrm{old}}\hat{\mathbf{z}}_{i}^{\mathrm
{old}}),\\
\mathrm{vec}(\hat{\bolds{\Xi}}^{\mathrm{new}}) &=&\Biggl[\sum
_{i=1}^{n}%
\biggl\{(\hat{\mathbf{V}}_{i}^{\mathrm{old}})^{-1}+\biggl(\frac{\nu
+m_{i}}{%
\nu+\hat{s}_{i}^{\mathrm{old}}}\biggr)\hat{\mathbf{z}}_{i}^{\mathrm{old}}(%
\hat{\mathbf{z}}_{i}^{\mathrm{old}})^{T}\biggr\}\otimes\mathbf
{B}_{i}^{T}%
\mathbf{B}_{i}\Biggr]^{-1} \\
&&{}\times\sum_{i=1}^{n}\biggl(\frac{\nu+m_{i}}{\nu+\hat
{s}_{i}^{\mathrm{old}%
}}\biggr)(\hat{\mathbf{z}}_{i}^{\mathrm{old}}\otimes\mathbf
{B}_{i}^{T})(%
\mathbf{x}_{i}\mathbf{-B}_{i}\hat{\bolds{\theta}}^{\mathrm{old}}),\\
(\hat{\sigma}^{2})^{\mathrm{new}} &=&\frac{1}{\sum
_{i=1}^{n}m_{i}}\Biggl[%
\sum_{i=1}^{n}\biggl(\frac{\nu+m_{i}}{\nu+\hat{s}_{i}^{\mathrm
{old}}}\biggr)%
\Vert\mathbf{x}_{i}-\mathbf{B}_{i}\hat{\bolds{\theta}}^{\mathrm
{old}}-%
\mathbf{B}_{i}\hat{\bolds{\Xi}}^{\mathrm{old}}\hat{\mathbf{z}}_{i}^{\mathrm{%
old}}\Vert^{2} \\
&&\hspace*{63pt}{}+\sum_{i=1}^{n}\operatorname{trace}\{\mathbf{B}_{i}\hat{\bolds{\Xi}}^{\mathrm{%
old}}(\hat{\mathbf{V}}_{i}^{\mathrm{old}})^{-1}(\hat{\bolds{\Xi}}^{\mathrm{%
old}})^{T}\mathbf{B}_{i}^{T}\}\Biggr],
\end{eqnarray*}
where $\hat{s}_{i}$ and $\hat{\mathbf{z}}_{i}$ are as before, and
$\mathbf{V}%
_{i}=\mathbf{I}_{d}+\bolds{\Xi}^{T}\mathbf{B}_{i}^{T}\mathbf
{B}_{i}\bolds{\Xi}/\sigma^{2}$. To obtain $\hat{\mathbf{H}}$ and $\hat{\bolds{\Lambda}}$
from $\hat{\bolds{\Xi}},$ we find the spectral decomposition of
$\hat{\bolds{\Xi}}^{T}\mathbf{J}\hat{\bolds{\Xi}}$, say, $\mathbf{UDU}^{T}$ with
$\mathbf{U}$
orthogonal and $\mathbf{D}$ diagonal, and set $\hat{\bolds{\Lambda}}=\mathbf{D}$ and
$\hat{\mathbf{H}}=\hat{\bolds{\Xi}}\mathbf{UD}^{-1/2}$.

As it is well known, the EM algorithm can take a large number of iterations
to converge; but for our estimators each iteration is very fast to compute.
Most of the computing time (in our Matlab implementation) is taken up
by the
recomputation of the spline basis matrix $\mathbf{B}_{i}$ for each $i$ on
each iteration, so the computing time grows mostly with $n$ and only
marginally with $d$, $p$ or the $m_{i}$s. To give an idea of the computing
times involved, each run of the EM algorithm for the simulated data in
Section \ref{sec:Simulations}, with $n=100$, takes approximately 15 seconds
on a common laptop computer with a 2.00GHz Intel Pentium processor.

In practice, the model dimension $d$ is not known a priori, so the
computation of the estimators is done in a sequential way. We recommend to
begin with a mean-only model ($d=0$), using $\hat{\bolds{\theta}}=\mathbf{0}
$ and $\hat{\sigma}^{2}=\sum_{i,j}x_{ij}^{2}/\sum_{i}m_{i}$ as initial
estimators for the EM iterations. Then proceed by adding one principal
component at a time, using the estimators of the previous $(d-1)$%
-dimensional model as initial estimators for the $d$-dimensional model. The~final dimension $d_{0}$ can be chosen subjectively or objectively.
Subjective approaches include choosing a $d$ that yields a small ratio
$\hat{%
\lambda}_{d}/\sum_{k=1}^{d}\hat{\lambda}_{k}$ or a small value of
$\hat{%
\lambda}_{d}$ compared to the noise variance $\hat{\sigma}^{2}$. Objective
model selection methods can be based on the maximization of the penalized
log-likelihood
%
\begin{eqnarray}
\sum_{i=1}^{n}\log f(\mathbf{x}_{i}|\hat{\bolds{\theta}},\hat{\mathbf{H}},\hat{\bolds\Lambda})-c_{n}\mathrm{df}, \label{eq:Penalized_likelihood}
\end{eqnarray}
where $f(\mathbf{x}_{i}|\hat{\bolds{\theta}},\hat{\mathbf{H}},\hat{\bolds\Lambda})$ is the
$t_{\nu}(\mathbf{B}_{i}\hat{\bolds{\theta}},\hat{\bolds\Sigma}_{i})$ density
evaluated at $\mathbf{x}_{i}$, $\mathrm{df}$ are the degrees of
freedom of
the model, and $c_{n}$ is a constant. Concretely, $c_{n}=1$ defines the AIC
criterion and $c_{n}=\log n/2$ the BIC criterion. This approach has been
used in the functional data context [Yao, M\"{u}ller and Wang (\citeyear{YaMuWa2005})] for normally
distributed data only, but Shen, Huang and Ye (\citeyear{ShHuYe2004}) justify their use for
exponential distributions in general. The~degrees of freedom of the model
are the number of parameters minus the number of orthonormality
restrictions. Another objective method that can be used, although in
practice it tends to underperform (\ref{eq:Penalized_likelihood}), is
cross-validation. Cross-validation would maximize
\[
\sum_{i=1}^{n}\log f\bigl(\mathbf{x}_{i}|\hat{\bolds{\theta}}_{(-i)},\hat{\mathbf{H}}_{(-i)},\hat{\bolds{\Lambda}}_{(-i)}\bigr),
\]
where $\hat{\bolds{\theta}}_{(-i)}$, $\hat{\mathbf{H}}_{(-i)}$
and $\hat{\bolds{\Lambda}}_{(-i)}$ are the estimators computed without
observation $%
\mathbf{x}_{i}$.

Another aspect that is rather subjective is the choice of basis functions
for $\mu(t)$ and $\phi_{k}(t)$, particularly the knot placement and
quantity. If a large number of knots is used, placement becomes less
important but regularization is necessary. This can be accomplished by
adding roughness penalty terms of the form $\alpha\int\{\mu^{\prime
\prime}(t)\}^{2}\,dt$ and $\alpha_{k}\int\{\phi_{k}^{\prime\prime
}(t)\}^{2}\,dt$ to the log-likelihood function (the resulting
modifications of
the EM algorithm are straightforward, since these terms are quadratic
in the
parameters). Selection of the smoothing parameters $\alpha$ and
$\alpha_{k}
$ can be done, again, either subjectively or objectively. The~penalized
log-likelihood approach, however, is not as straightforward to
implement as
before, because the degrees of freedom of the model are not as easy to
calculate when the fitted values $\hat{\mathbf{x}}_{i}$ are not linear
functions of the data [Ye (\citeyear{Ye1998}); Efron (\citeyear{Ef2004})]. Cross-validation, on the
other hand, can be implemented as easily as usual despite its shortcomings.
Nevertheless, since $\bolds{\theta}$ and the $\bolds{\eta}_{k}$s are
model parameters common to all curves, the estimators $\hat{\bolds{\theta}}$
and $\{\hat{\bolds{\eta}}_{k}\}$ ``borrow
strength'' across individuals and then the choice of
smoothing parameters is less problematic than if each curve were smoothed
individually. This is based on our experience with spline smoothing rather
than on formal mathematical results, although for kernel smoothers Yao, M\"{u}ller and Wang
(\citeyear{YaMuWa2005}) and Kneip (\citeyear{Kn1994}) have indeed established rates of
convergence of
the estimators and the bandwidths that depend fundamentally on the
number of
curves $n$ rather than the number of observations per curve $m_{i}$.

\section{Asymptotic properties}\label{sec:Properties}

The~distributional assumptions made in Section \ref{sec:t-estimators} were
just working assumptions to derive robust estimators of $\mu(t)$ and
the $%
\phi_{k}(t)$s. In this section we will study the consistency of the
estimators under broader conditions. We will also study their
sensitivity to
outliers, as quantified by the influence function.

To simplify, let us assume that the individual time grids $\mathbf{t}%
_{1},\ldots,\mathbf{t}_{n}$ are i.i.d.~realizations of a random
vector $%
\mathbf{t}\in\mathbb{R}^{m}$, so $m_{i}=m$ for all $i$. Let $\mathbf
{w}=(%
\mathbf{t},\mathbf{x})$ and let us collect all model parameters in a single
vector $\bolds{\xi}=(\bolds{\theta},\bolds{\eta}_{1},\ldots
,\bolds\eta%
_{d},\lambda_{1},\ldots,\lambda_{d},\sigma^{2})$. The~estimating
equations (\ref{eq:estim_eq_1}) to (\ref{eq:estim_eq_4}) can be
expressed as
a single system of equations $\sum_{i=1}^{n}\bolds{\psi}(\mathbf
{w}_{i},%
\hat{\bolds{\xi}})=\mathbf{0}$ for an appropriate function
$\bolds{\psi}(%
\mathbf{w},\cdot)\dvtx \mathbb{R}^{(p+1)(d+1)}\rightarrow\mathbb{R}^{(p+1)(d+1)}
$. Estimators of this type are called $M$-estimators, or sometimes \mbox{$Z$-estimators} [Maronna, Martin and Yohai (\citeyear{MaMa2006}), Chapter 3; Van der
Vaart (\citeyear{Va1998}), Chapter~5]. For such estimators the notion of Fisher
consistency is
useful. Suppose $\mathbf{w}=(\mathbf{t},\mathbf{x})$ follows model
(\ref%
{eq:raw-data-model}) with parameter $\bolds{\xi}=\bolds{\xi}_{0}$, and let $F_{0}$
be the resulting distribution of $\mathbf{w}$. Let $\bolds{\xi}=\bolds{\xi}(F_{0})
$ be the solution to the eqnarray $\mathrm{E}_{F_{0}}\{\bolds{\psi}(\mathbf{w},\bolds\xi)%
\}=\mathbf{0}$. In principle, $\bolds{\xi}(F_{0})$ need not be
equal to
the true model parameter $\bolds{\xi}_{0}$; if it is, the estimator is
said to be Fisher consistent [Maronna, Martin and Yohai (\citeyear{MaMa2006}), page~67]. It turns out
that under some regularity conditions, $M$-estimators $\hat{\bolds{\xi}}$
converge in probability to $\bolds{\xi}(F_{0})$ as $n$ goes to infinity;
then, under those regularity conditions, Fisher consistency implies the
usual consistency [Van der Vaart (\citeyear{Va1998}), Theorem~5.9].

The~next theorem shows that $\hat{\bolds{\theta}}$ and the $\hat{\bolds{\eta}}_{k}$s are Fisher consistent under broad conditions, whereas
$\hat{%
\sigma}^{2}$ and the $\hat{\lambda}_{k}$s are off by a common factor [this
is typical of $M$-estimators of scale parameters; see Maronna, Martin and Yohai (\citeyear{MaMa2006}),
Chapter~6.12].

\begin{theorem}
\label{theo:Fisher-consistency}If $\mathbf{w}=(\mathbf{t},\mathbf{x})$ follows model
(\ref%
{eq:raw-data-model}) with parameter $\bolds{\xi}_{0}$, and $\mathbf{(z,\bolds\varepsilon)}$
has a joint spherical distribution, then $\bolds{\xi}%
(F_{0})=(\bolds{\theta}_{0},\bolds{\eta}_{01},\ldots,\bolds{\eta}%
_{0d},\beta_{0}\lambda_{01},\break\ldots,\beta_{0}\lambda_{0d}, \beta
_{0}\sigma_{0}^{2})$ with $\beta_{0}>0$ a factor that depends on the
distribution of $(\mathbf{z},\bolds\varepsilon)$ and on $\nu$ but not on the
model parameter $\bolds{\xi}_{0}$.
\end{theorem}

Theorem \ref{theo:Fisher-consistency} implies that the estimators of
$\mu$
and the $\phi_{k}$s derived under a $t_{\nu}$ distributional
assumption on
$(\mathbf{z},\bolds\varepsilon)$ are actually Fisher consistent under \textit{any}
spherical distribution of $(\mathbf{z},\bolds\varepsilon)$, including the Normal
distribution or a $t_{\nu^{\ast}}$ distribution with $\nu^{\ast
}\neq\nu
$. The~estimators of $\sigma^{2}$ and the $\lambda_{k}$s, although not
Fisher-consistent, are off by a common factor $\beta_{0}$, which implies
that the ratios $\hat{\lambda}_{d}/\hat{\sigma}^{2}$ and $\hat
{\lambda}%
_{d}/\sum_{k=1}^{d}\hat{\lambda}_{k}$ are Fisher-consistent.

Now we turn our discussion to the outlier sensitivity of the estimators.
Outlier sensitivity can be measured by the influence function [Maronna, Martin and Yohai (\citeyear{MaMa2006}), Chapter~3], which is defined as
\begin{eqnarray*}
\operatorname{IF}(\mathbf{w};\hat{\bolds{\xi}},F_{0})=\lim
_{\varepsilon\searrow
0}\frac{1}{\varepsilon}\bigl\{\bolds{\xi}\bigl((1-\varepsilon
)F_{0}+\varepsilon
\delta_{\mathbf{w}}\bigr)-\bolds{\xi}(F_{0})\bigr\},
\end{eqnarray*}
where $\delta_{\mathbf{w}}$ is the point-mass distribution at
$\mathbf{w}$.
The~gross-error sensitivity of $\hat{\bolds{\xi}}$ is defined as
$\gamma
^{\ast}=\sup_{\mathbf{w}}\Vert\operatorname{IF}(\mathbf{w};\hat{\bolds{\xi}}%
,F_{0})\Vert$. Note that for a small contamination proportion~$\varepsilon$, the asymptotic bias caused by $\delta_{\mathbf{w}}$ is
approximately $%
\varepsilon\operatorname{IF}(\mathbf{w};\hat{\bolds{\xi}},F_{0})$.
Therefore, if
$\gamma^{\ast}<\infty$, the bias is bounded regardless of the
location of
the outliers and the estimator $\hat{\bolds{\xi}}$ is said to be locally
robust.

For regular $M$-estimators, it can be shown that $\operatorname{IF}(\mathbf
{w};%
\hat{\bolds{\xi}},F_{0})=-\mathbf{M}^{-1}\bolds{\psi}(\mathbf
{w},\break \bolds{\xi}(F_{0}))$, where $\mathbf{M}=\mathrm{E}_{F_{0}}\{
\partial
\bolds{\psi}(\mathbf{w},\bolds\xi)/\partial\bolds{\xi}%
^{T}\vert_{\bolds{\xi}=\bolds{\xi}(F_{0})}\}$ [Maronna, Martin and Yohai (\citeyear{MaMa2006}),
Chapter~3]. Then $\gamma^{\ast}\leq\lambda_{\min}^{-1/2}(\mathbf
{MM}%
^{T})\sup_{\mathbf{w}}\Vert\bolds{\psi}(\mathbf{w},\bolds\xi
(F_{0}))\Vert$,
where\break $\lambda_{\min}(\mathbf{A})$ denotes the smallest eigenvalue
of $%
\mathbf{A}$, so $\gamma^{\ast}<\infty$ as long as $\mathbf{M}$ is
invertible and $\bolds{\psi}(\mathbf{w},\bolds\xi)$ is bounded in $\mathbf{w}$.
This is
true for our estimating functions $\bolds{\psi}$, so the \mbox{$t$-model}
estimators $\hat{\bolds{\xi}}$ are locally robust.

Influence functions are also useful for the computation of asymptotic
variances. Under appropriate regularity conditions, $\sqrt{n}(\hat{\bolds{\xi}}-\bolds{\xi}(F_{0}))$ converges in distribution to a $N(\mathbf
{0},\mathbf{V})$
with $\mathbf{V}=\mathrm{E}\{\operatorname{IF}(\mathbf{w};\hat{\bolds{\xi}},F_{0})%
\operatorname{IF}(\mathbf{w};\hat{\bolds{\xi}},F_{0})^{T}\}$ [Van der Vaart
(\citeyear{Va1998}), Theorem 5.21]. This result is useful, for instance, to derive
asymptotic confidence bands for $\mu(t)$, provided one can obtain a more
explicit expression for the $p\times p$ block of $\mathbf{V}$ that
corresponds to the asymptotic variance of $\hat{\bolds{\theta}}$.
In some
cases this is possible, as the next theorem shows.

\begin{theorem}
\label{theo:IF}
If $\mathbf{w}=(\mathbf{t},\mathbf{x})$ follows model (\ref{eq:raw-data-model})
and $(\mathbf{z},\bolds\varepsilon)$ has a joint spherical distribution,
then $%
\mathbf{M}$ has a block structure
\begin{eqnarray*}
\mathbf{M}=\left[
\matrix{
\mathbf{M}_{11} & \mathbf{0} \cr
\mathbf{0} & \mathbf{M}_{22}%
}
\right]
\end{eqnarray*}
with $\mathbf{M}_{11}\in\mathbb{R}^{p\times p}$ given by
\begin{eqnarray*}
\mathbf{M}_{11}=\mathrm{E}_{F_{0}}\biggl\{g(\mathbf{w})\mathbf
{B}^{T}(\mathbf{%
t})\frac{1}{\beta_{0}}\bolds{\Sigma}^{-1}(\mathbf{t})\mathbf
{B}(\mathbf{t}%
)\biggr\},
\end{eqnarray*}
where
\begin{eqnarray*}
g(\mathbf{w})=\frac{2}{m}\frac{(\nu+m)s(\mathbf{w})/\beta_{0}}{\{
\nu+s(%
\mathbf{w})/\beta_{0}\}^{2}}-\frac{\nu+m}{\nu+s(\mathbf{w})/\beta_{0}},
\end{eqnarray*}
$\mathbf{B}(\mathbf{t})=[b_{k}(t_{j})]_{(j,k)}$, $\bolds{\Sigma
}(\mathbf{t}%
)=\mathbf{B}(\mathbf{t})\mathbf{H}_{0}\bolds{\Lambda}_{0}\mathbf
{H}_{0}^{T}%
\mathbf{B}^{T}(\mathbf{t})+\sigma_{0}^{2}\mathbf{I}_{m}$,
$s(\mathbf{w})=\{%
\mathbf{x}-\bolds{\mu}_{0}(\mathbf{t})\}^{T}\times \break \bolds{\Sigma
}^{-1}(\mathbf{t%
})\{\mathbf{x}-\bolds{\mu}_{0}(\mathbf{t})\}$ and $\bolds{\mu
}_{0}(%
\mathbf{t})=\mathbf{B}(\mathbf{t})\bolds{\theta}_{0}$. Furthermore,
\begin{eqnarray*}
\operatorname{IF}(\mathbf{w};\hat{\bolds{\theta}},F_{0})=-\mathbf
{M}_{11}^{-1}%
\biggl(\frac{\nu+m}{\nu+s(\mathbf{w})/\beta_{0}}\biggr)\mathbf
{B}^{T}(%
\mathbf{t})\frac{1}{\beta_{0}}\bolds{\Sigma}^{-1}(\mathbf{t})\{
\mathbf{x}-%
\bolds{\mu}_{0}(\mathbf{t})\}.
\end{eqnarray*}
\end{theorem}

From Theorem \ref{theo:IF} we see that the asymptotic covariance
matrix of $%
\hat{\bolds{\theta}}$ has the form $\mathbf{M}_{11}^{-1}\mathbf
{AM}%
_{11}^{-1}$, and due to the block structure of $\mathbf{M}$, $\hat{\bolds{%
\theta}}$ is asymptotically independent of $\{\hat{\bolds{\eta
}}_{k}\}$, $\{%
\hat{\lambda}_{k}\}$ and $\hat{\sigma}^{2}$. The~matrices $\mathbf{M}_{11}$
and $\mathbf{A}$ can be estimated by
\begin{eqnarray*}
\hat{\mathbf{M}}_{11}=\frac{1}{n}\sum_{i=1}^{n}\hat{g}_{i}\mathbf
{B}_{i}^{T}%
\hat{\bolds{\Sigma}}_{i}^{-1}\mathbf{B}_{i}
\end{eqnarray*}
and
\begin{eqnarray*}
\hat{\mathbf{A}}=\frac{1}{n}\sum_{i=1}^{n}\biggl(\frac{\nu+m}{\nu
+\hat{s}%
_{i}}\biggr)^{2}\mathbf{B}_{i}^{T}\hat{\bolds{\Sigma
}}_{i}^{-1}(\mathbf{x}%
_{i}-\hat{\bolds{\mu}})(\mathbf{x}_{i}-\hat{\bolds{\mu
}})^{T}\hat{\bolds{%
\Sigma}}_{i}^{-1}\mathbf{B}_{i},
\end{eqnarray*}
where
\begin{eqnarray*}
\hat{g}_{i}=\frac{2(\nu+m)\hat{s}_{i}}{m(\nu+\hat
{s}_{i})^{2}}-\frac{\nu+m%
}{\nu+\hat{s}_{i}}.
\end{eqnarray*}
Note that, by Theorem \ref{theo:Fisher-consistency}, $\hat{\bolds{\Sigma}}%
_{i}$ and $\hat{s}_{i}$ are consistent estimators of $s(\mathbf{w}%
_{i})/\beta_{0}$ and $\bolds{\Sigma}^{-1}(\mathbf{t}_{i})/\beta
_{0}$, so
$\hat{\mathbf{M}}_{11}^{-1}\hat{\mathbf{A}}\hat{\mathbf{M}}_{11}^{-1}$ is a
consistent estimator of $\mathbf{M}_{11}^{-1}\mathbf{AM}_{11}^{-1}$.

\section{Simulation study}\label{sec:Simulations}

\subsection{Assessment of parameter estimators}

We studied the finite-sample behavior of the estimators by simulation. We
were mainly interested in the relative efficiency of the estimators for
normally distributed data and in their bias under outlier contamination.
Three estimators were considered: the maximum likelihood estimator for
(a) the Normal model [James, Hastie and Sugar (\citeyear{JaHaSu2000})], (b) the Cauchy model,
which is a \mbox{$t$-model} with $\nu=1$, and (c) the \mbox{$t$-model} with
$\nu
=5 $. As spline basis we chose cubic splines with five equidistant
knots. We
considered different simulation scenarios (described below) but only
part of
the results are reported here (Table \ref{tab:simulation-results}).
The~rest
can be found in the Technical Report.

To assess the efficiency of the estimators, we simulated data from the
two-component model
%
\begin{eqnarray}
x_{ij}=\mu(t_{ij})+\sum_{k=1}^{2}z_{ik}\sqrt{\lambda_{k}}\phi
_{k}(t_{ij})+\sigma\varepsilon_{ij}, \label{eq:simulated_model}
\end{eqnarray}
with $\mu(t)=0$ and $\phi_{k}(t)=\sqrt{2}\sin(k\pi t)$, for $t\in
\lbrack
0,1]$. The~component scores $z_{ik}$ and the random errors $\varepsilon
_{ij} $ were independent $N(0,1)$ and $\lambda_{1}=1$, $\lambda
_{2}=0.5$, $%
\sigma^{2}=0.25$. Three designs were considered for the $t_{ij}$s:
(i)
$m=20$ fixed uniformly spaced points in $[0,1]$, (ii) $m=20$ random
points (which vary from curve to curve) with uniform distribution in
$[0,1]$%
, and (iii) $m_{i}$ random points with uniform distribution in
$[0,1]$%
, where $m_{1},\ldots,m_{n}$ was a sample from a Poisson random variable
with mean 15. The~third design is the one that best resembles sparse and
irregularly observed data. As sample sizes we took $n=50$, $n=100$ and $
n=200 $. Each sampling situation was replicated 500 times. Root mean squares
of $\Vert\hat{\mu}-\mu\Vert$ and $\Vert\hat{\phi}_{1}-\phi
_{1}\Vert$
are given in Table \ref{tab:simulation-results}, for grid design (ii)
and sample size $n=100$. The~relative behavior of the estimators is similar
for the other designs and sample sizes, as can be seen in the more detailed
results shown in the Technical Report. We see that the $t$-model estimators
are generally less efficient than the Normal-model estimators, as expected,
but the loss of efficiency is minimal. We also note that the estimators
$%
\hat{\mu}$ were obtained by fitting a mean-only model to the
simulated data,
whereas the estimators $\hat{\phi}_{1}$ where obtained by fitting a
one-component model to the data; therefore, the models were always
underspecified, but this did not seem to affect the consistency of the
estimators (the boxplots in the Technical Report show that the errors
decrease as $n$ increases).

\begin{table}
\caption{Simulation results. Root mean squared errors of different
estimators for noncontaminated normal data and outlier-contaminated
data}\label{tab:simulation-results}
\begin{tabular*}{\textwidth}{@{\extracolsep{\fill}}lcccccccccccc@{}}
\hline
& & & & \multirow{2}{32pt}[-5pt]{\centering{\textbf{No contam.}}} & & \multicolumn{3}{c}{\textbf{Endogenous contam.}} & &\multicolumn{3}{c@{}}{\textbf{Exogenous contam.}}\\[-6pt]
& & & &  & & \multicolumn{3}{c}{\hrulefill} & &\multicolumn{3}{c@{}}{\hrulefill}\\
\textbf{Estim.} & & \textbf{Model} & & & & \textbf{10\%} & \textbf{20\%} & \textbf{30\%} & & \textbf{10\%} & \textbf{20\%} &\textbf{30\%} \\
\hline
$\hat{\mu}$ & & Normal & & 0.142 & & 0.427 & 0.819 & 1.205 & & 0.367 & 0.703& 1.040 \\
& & Cauchy & & 0.169 & & 0.190 & 0.247 & 0.330 & & 0.162 & 0.184 & 0.212 \\
& & $t_{5}$ & & 0.159 & & 0.183 & 0.254 & 0.365 & & 0.153 & 0.179 & 0.224 \\
\hline
$\hat{\phi}_{1}$ & & Normal & & 0.142 & & 1.091 & 1.331 & 1.363 & & 0.942& 1.265 & 1.290 \\
& & Cauchy & & 0.165 & & 0.299 & 0.627 & 1.006 & & 0.158 & 0.189 & 0.220 \\
& & $t_{5}$ & & 0.163 & & 0.338 & 0.673 & 1.087 & & 0.152 & 0.183 & 0.232 \\
\hline
\end{tabular*}
\end{table}

To assess the robustness of the estimators, two types of outliers were
considered; we call them endogenous and exogenous. Endogenous outliers are
curves that belong to the space spanned by $\{\phi_{1},\phi_{2}\}$ just
like the rest of the data, only that the component scores $\mathbf{z}_{i}$
follow a different distribution. Exogenous outliers, on the contrary, are
curves that do not belong to the space spanned by $\{\phi_{1},\phi
_{2}\}$.
In these simulations we generated exogenous outliers by taking linear
combinations of $\phi_{1}$, $\phi_{2}$ and $\phi
_{3}(t)=c\{t(1-t)\}^{1/2}\sin\{2\pi(1+2^{(9-4k)/5})/(t+2^{(9-4k)/5})\}$
with $k=5$ (the so-called ``Doppler
function,'' with $c$ such that $\Vert\phi_{3}\Vert=1$).
Three contamination proportions were considered for the scenarios described
below: $\varepsilon=0.10$, $\varepsilon=0.20$ and $\varepsilon=0.30$. The~time grid was generated following the uniform random design (ii), and
the sample size was $n=100$. Each scenario was replicated 500 times.

Let us first examine the robustness of $\hat{\mu}$. Endogenous
outliers were
generated by replacing $\varepsilon n$ component scores $z_{i1}$ with a
large constant $K$ (so the outlying curves were virtually identical to
$K%
\sqrt{\lambda_{1}}\phi_{1}$), whereas exogenous outliers were
generated by
adding $K\sqrt{\lambda_{1}}\phi_{3}$ to $\varepsilon n$ sample
curves. We
considered two contaminating constants, $K=4$ and $K=8$, but only the
results for $K=4$ are reported in Table \ref{tab:simulation-results} (the
results for $K=8$ are given in the Technical Report). Since the
``true'' mean for these samples are $%
\varepsilon K\phi_{1}$ and $\varepsilon K\phi_{3}$, respectively,
because $%
\lambda_{1}=1$, the root mean squared errors should be approximately $%
\varepsilon K$ for nonrobust estimators. This is exactly what we see in
Table \ref{tab:simulation-results} for the Normal-model estimator. In
contrast, \mbox{$t$-model} estimators show remarkably low biases, even for
contamination proportions as high as 30\%.

To study the robustness of $\hat{\phi}_{1}$, endogenous outliers were
generated by replacing $\varepsilon n/2$ scores $z_{i2}$ with $K\sqrt{%
\lambda_{2}}$ and $\varepsilon n/2$ with $-K\sqrt{\lambda_{2}}$; exogenous
outliers were generated by adding $K\sqrt{\lambda_{1}}\phi_{3}$ to $%
\varepsilon n/2$ sample curves and subtracting the same quantity to
other $%
\varepsilon n/2$ sample curves. As before, we used $K=4$ and $K=8$ but only
report the case $K=4$ here, since the other results are similar. Note that
these symmetric contaminations affect $\hat{\phi}_{1}$ but do not
affect $%
\hat{\mu}$, because they alter the covariance structure without
changing the
mean. In fact, the endogenously contaminated data follows model (\ref%
{eq:simulated_model}) with $\lambda_{1}^{\ast}=(1-\varepsilon
)\lambda_{1}
$ and $\lambda_{2}^{\ast}=(1-\varepsilon)\lambda_{2}+\varepsilon
\lambda
_{2}K^{2}$, so $\lambda_{2}^{\ast}$ can be actually larger than
$\lambda
_{1}^{\ast}$ if $K$ is big enough, in which case we expect the root mean
squared error of a nonrobust estimator to be close to $\Vert\phi
_{2}-\phi
_{1}\Vert=\sqrt{2}$. This is what we observe in Table \ref%
{tab:simulation-results}. The~exogenously contaminated data also follows
model (\ref{eq:simulated_model}) with three components, but the components
are not $\phi_{1}$, $\phi_{2}$ and~$\phi_{3}$ (because they are not
orthogonal). We see in Table \ref{tab:simulation-results} that endogenous
outliers have a more deleterious effect on the estimators than exogenous
outliers. In fact, the \mbox{$t$-model} estimators are practically unaffected by
exogenous outliers. Under endogenous contaminations, the performance of
the $%
t$-model estimators deteriorates for large contamination proportions,
although they still outperform the Normal-model estimators.

Overall, the conclusion from this Monte Carlo study is that $t$-model
estimators are highly resistant to outliers, even for relatively large
contamination proportions, and have a high relative efficiency for Normal
data. Given that their computational complexity is comparable to that of
Normal-model estimators, we think that they are a practical and safer
alternative. In particular, we recommend the use of Cauchy-model estimators,
since they are the most robust in the $t$ family and are not much less
efficient than $t_{5}$-model estimators for Normal data.\looseness=1

\subsection{Assessment of model selection criteria}

\begin{table}[b]
\caption{Simulation results. Percentage of times AIC and BIC select a
two-component model and
a three-component model, for Normal and Cauchy estimators and several
contamination proportions}\label{tab:simulation-results-model-selection}
\begin{tabular*}{\textwidth}{@{\extracolsep{4in minus 4in}}lccccc@{}}
\hline
& &  \multicolumn{4}{c@{}}{\textbf{Contamination proportion}} \\[-6pt]
& &  \multicolumn{4}{c@{}}{\hrulefill} \\
$\bolds{n}$ &  \textbf{Method} &  \multicolumn{1}{c}{\textbf{0\%}} & \multicolumn{1}{c}{\textbf{10\%}} & \multicolumn{1}{c}{\textbf{20\%}} & \multicolumn{1}{c@{}}{\textbf{30\%}} \\
\hline
$20$ &  AIC-Nor &  $(99,1)$ & $(1,74)$ & $(3,65)$ & $(6.3,62.3)$ \\
& BIC-Nor & $(99.7,0.3)$ & $(1,79.3)$ & $(4.7,71.3)$ & $(11.3,66.3)$
\\
&  AIC-Cau &  $(98,2)$ & $(74,24.3)$ & $(12.3,83.7)$ & $(0,87.3)$ \\
&  BIC-Cau &  $(99.7,0.3)$ & $(84.7,15.3)$ & $(20.7,78)$ & $(0.3,94.7)$
\\ [5pt]
$60$ &  AIC-Nor &  $(100,0)$ & $(0.3,60.7)$ & $(0.3,53.7)$ & $(1.3,62)$
\\
&  BIC-Nor &  $(100,0)$ & $(0.3,62.3)$ & $(0.3,55.3)$ & $(2,66)$ \\
&  AIC-Cau & $(100,0)$ & $(79.3,20)$ & $(0.3,76)$ & $(0,67.7)$ \\
&  BIC-Cau &  $(100,0)$ & $(89.3,10.7)$ & $(1,94.7)$ & $(0,92)$ \\
\hline
\end{tabular*}
\end{table}

We also ran a Monte Carlo study to evaluate the performance of the AIC and
BIC criteria for selection of the model dimension $d$. We generated data
from the two-component model (\ref{eq:simulated_model}) and from a
symmetrically contaminated model with exogenous outliers, as explained
above. Since exogenous contamination introduces a spurious third direction
of variability, the expected effect on the AIC and BIC criteria is an
overestimation of the model dimension.

We compared two types of estimators: the Normal-model estimators and the
Cauchy-model estimators. As before, we chose cubic splines with five
equidistant knots as spline basis; then $p=9$ and, for the true model
with $%
d=2$, the degrees of freedom are 27. We considered two sample sizes, $n=20$
and $n=60$ (note that in the former case $n$ is less than the degrees of
freedom of the true model). The~models were fitted in a sequential way, as
suggested in Section \ref{sec:t-estimators}, from $d=0$ to $d=4$. Each
sampling situation was replicated 300 times.

The~results are summarized in Table \ref%
{tab:simulation-results-model-selection}. We show two outputs: the
percentage of the samples for which the right model is selected and the
percentage of the samples for which the next model ($d=3$) is selected; the
remaining percentage would correspond to the four-dimensional model, since
we observed that models with $d<2$ were never selected. For noncontaminated
data, it is clear that the criteria have no trouble selecting the right
model, neither for Normal nor for Cauchy estimators. For low contamination
levels ($\varepsilon=0.10$) the AIC and BIC based on Cauchy estimators
select the right model in the vast majority of cases, with the BIC criterion
being clearly superior; the nonrobust Normal estimators, in contrast,
almost never led to the right choice of model. For larger contamination
proportions, even the robust estimators break down; but even then we note
that the BIC based on Cauchy-model estimators outperforms the alternatives,
since it selects the slightly overspecified model $d=3$ most of the
time and
very rarely leads to the worst choice $d=4$, in contrast to the other
methods. All things considered, we think the BIC based on Cauchy estimators
is a recommendable criterion for selection of the number of components.

\eject
\section{Examples}\label{sec:Example}

\subsection{Internet traffic}

Accurate modeling of Internet traffic data is essential for an efficient
allocation of computational resources. In this section we show that
just a
couple of atypical curves can lead to seriously misleading results. The~data, previously analyzed by Zhang et al.~(\citeyear{ZhMaShZh2007}), was collected at the main
Internet link of the University of North Carolina during seven consecutive
weeks (from June 9 to July 25, 2003). The~traffic is measured in packet
counts, every half hour. The~logarithm of the data for week days is
shown in
Figure~\ref{fig:traffic_data}(a) and for weekend days in Figure~\ref%
{fig:traffic_data}(b).

Although the data is very noisy, we see that the trajectories follow a
regular pattern, which is different for week days than for weekends.
Here we
analyze only the 35 week days. There is a very clear outlier in
Figure~\ref%
{fig:traffic_data}(a), a curve that actually looks like a weekend
trajectory. This curve corresponds to the Fourth of July. A~more subtle
atypical curve corresponds to June 27, the second day of classes and the
last day for late registration for the Second Summer Session. That day the
traffic peaked two hours earlier than usual, and also decreased earlier than
usual in the afternoon.

\begin{figure}

\includegraphics{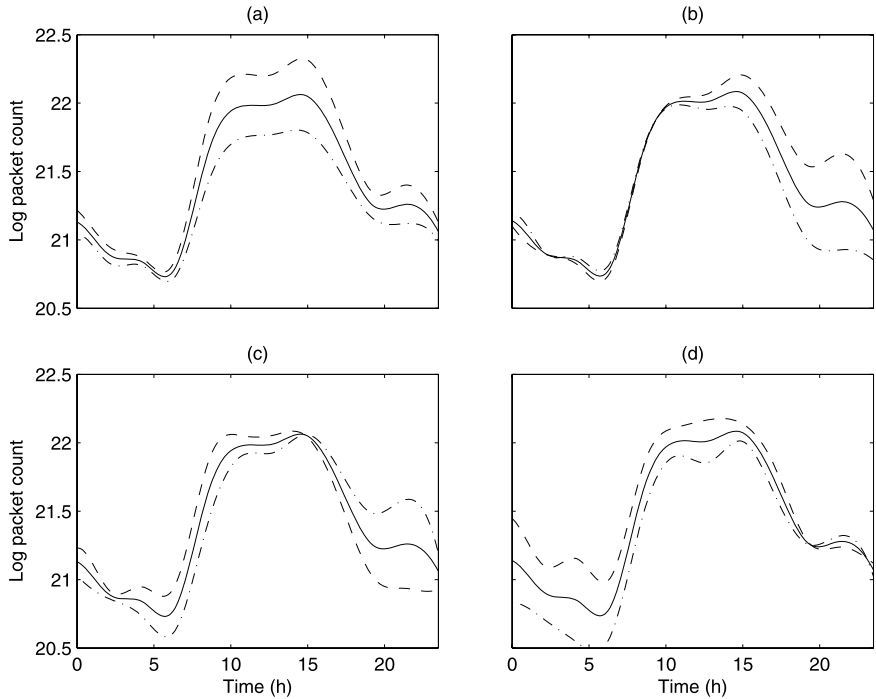}

\caption{Internet traffic data.
Estimators of the mean (-----) and the mean plus ($---$) and minus
($-\cdot
- $) a constant times the principal component, for Normal-model estimators
[(\textup{a}), (\textup{c})] and Cauchy-model
estimators [(\textup{b}), (\textup{d})] of the first [(\textup{a}), (\textup{b})] and
second [(\textup{c}), (\textup{d})] principal components.}\label{fig:Mean_and_pcs}
\end{figure}

\begin{figure}

\includegraphics{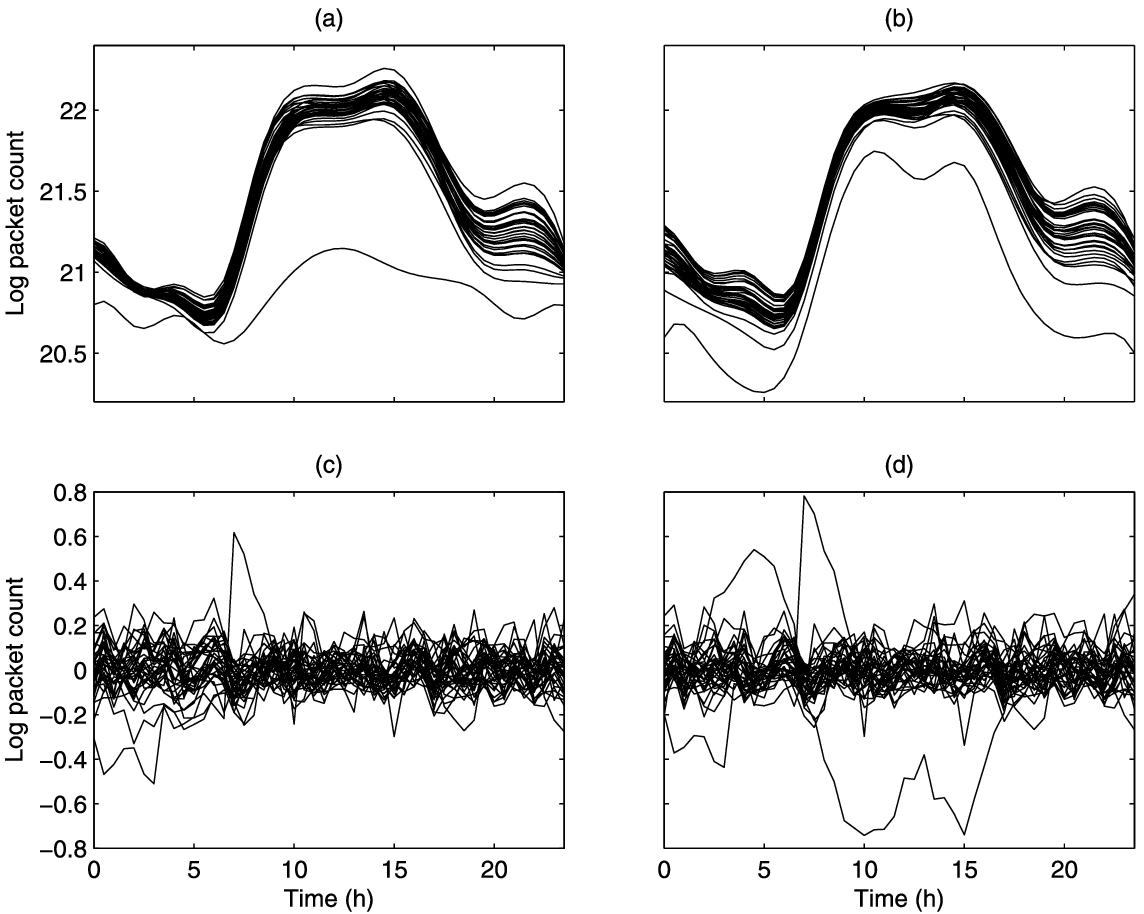}

\caption{Internet traffic data. Estimated
trajectories [(\textup{a}), (\textup{b})] and residuals [(\textup{c}),
(\textup{d})] from Normal-model estimators
[(\textup{a}), (\textup{c})] and Cauchy-model
estimators [(\textup{b}), (\textup{d})].}\label{fig:Fits_and_errors}
\end{figure}

We estimated the mean and the first two principal components using
Normal-model and Cauchy-model estimators based on cubic splines with 10
equispaced knots. The~results are shown in Figure~\ref{fig:Mean_and_pcs}.
Rather than plotting the principal components themselves, we show their
effect on the mean, by plotting $\hat{\mu}$ plus/minus a constant
times $%
\hat{\phi}_{k}$. This makes interpretation easier. We see that the mean
estimators are similar but the principal components are completely
different. The~Normal-model estimator of the first component is an amplitude
effect (above/below the mean, but parallel to it) and the second component
is a shape component (traffic higher than the mean until 3 p.m. and lower
than the mean afterward). The~first component is clearly dominant,
since $%
\hat{\lambda}_{1}=0.329$ and $\hat{\lambda}_{2}=0.085$. It is very suspicious
that these components essentially mimic the two outliers; in fact, July 4
has the largest first component score and June 27 the largest second
component score.

On the other hand, the Cauchy-model estimators of the components explain
amplitude variability at the end of the day (the first component) and
at the
beginning of the day (the second component), with the total variability
roughly equally split ($\hat{\lambda}_{1}=0.176$ and $\hat{\lambda
}_{2}=0.123$%
). Of course, the fact that the two methods produce different estimators
does not automatically imply that the Cauchy-model estimators are better,
but a residual analysis confirms this. Cauchy-model estimators produce
smaller residual norms $\Vert\mathbf{x}_{i}-\hat{\mathbf{x}}_{i}\Vert$
than Normal-model estimators for 25 of the 35 observations. The~median
residual norm for the Cauchy fit is $0.556,$ while for the Normal fit it
is $0.592$. Figure \ref{fig:Fits_and_errors} shows individual predictors and
residuals; undoubtedly, Cauchy-model estimators offer an overall better fit
(except for the Fourth of July outlier). Normal-model estimators show a
particularly poor fit for the Internet traffic between 0 and 6 a.m.

\subsection{Child diabetes study}

\begin{figure}

\includegraphics{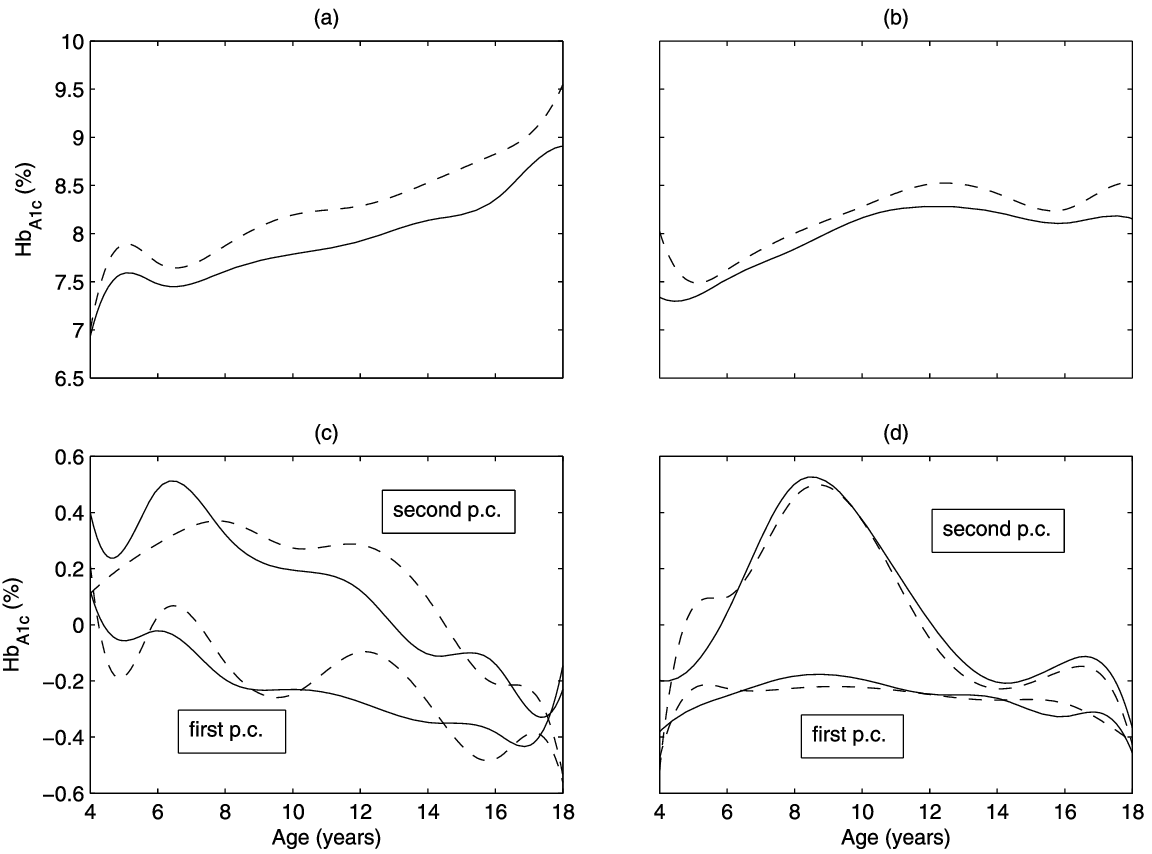}

\caption{Child diabetes data. Estimated
means [(\textup{a}), (\textup{b})] and leading principal components [(\textup{c}), (\textup{d})] of $Hb_{A1c}$
trajectories for females [(\textup{a}), (\textup{c})] and
males [(\textup{b}), (\textup{d})], using Normal (dashed
line) and Cauchy (solid line) maximum likelihood estimators.}\label{fig:MeanPC}
\end{figure}

Glycated hemoglobin ($Hb_{A1c}$) levels are often used as a measure of
average plasma glucose concentration over certain periods of time.
Figure %
\ref{fig:trajectories} shows trajectories of $Hb_{A1c}$ levels for diabetic
children who underwent treatment at the Children's Hospital of the
University of Zurich. The~profiles are very irregularly sampled and noisy.
For girls, the minimum number of observations per trajectory is 2, the
median 33 and the maximum 55; for boys, the minimum number of observations
per trajectory is 2, the median 33 and the maximum 56. For such irregular
data individual smoothing is impractical, even impossible for the shortest
trajectories. The~presence of at least one outlying curve is plain to
see in
Figure~\ref{fig:trajectories}(a), although it is hard to tell by visual
inspection if there are any other outliers.

We estimated the mean and the principal components for each sex, using
Normal and Cauchy maximum likelihood estimators. Cubic splines with 6
equispaced knots were used as basis functions. The~BIC based on Normal-model
estimators selects a three-component model for both sexes, while the BIC
based on Cauchy-model estimators selects a four-component model;
however, in
the latter case $\hat{\lambda}_{4}$ is very small compared to $\hat
{\sigma}%
^{2}$ and the other $\hat{\lambda}$s, so we settled for a three-component
model. Figure \ref{fig:MeanPC} shows the estimators of the mean and
the two
leading components (the third one is omitted for better visibility). For
males, both methods produce similar estimators, but for females the
differences are striking. The~Normal-model estimators not only overestimate
the mean but also provide a very irregular estimator of the first principal
component; the estimator of the second component is also substantially
different from the Cauchy-model estimator.

\begin{figure}

\includegraphics{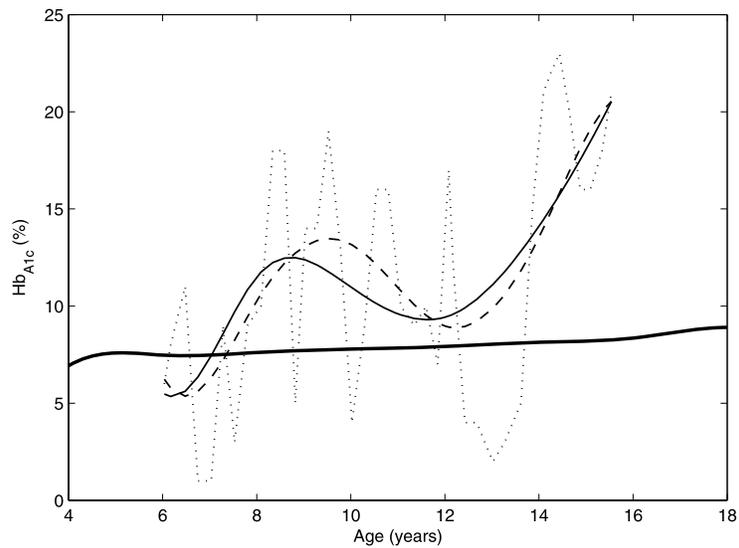}

\caption{Child diabetes data. Outlying
trajectory (dotted line), Cauchy-model estimator of the mean (thick solid
line), and fitted trajectory using Cauchy-model predictor (thin solid line)
and Normal-model predictor (dashed line).}\label{fig:Outliers}
\end{figure}

The~trajectory with the largest mean squared residual for girls is
shown in
Figure~\ref{fig:Outliers}. This is a patient whose diabetes level was clearly
out of control. We see that the Normal-model estimator provides a somewhat
better fit for this curve than the Cauchy estimator, but this is at the
expense of a poorer fit for the rest of the individuals. The~mean squared
residual of this observation is 22.6 for the Normal fit and 24.8 for the
Cauchy fit. However, the three quartiles of the mean squared residuals for
the whole sample are 0.18, 0.40 and 0.63 for the Cauchy fit, and 0.24, 0.43
and 0.71 for the Normal fit, so the Cauchy fit is better overall. Another
confirmation of this is that the Normal-model estimators obtained after
eliminating the outlying trajectory are very similar to the Cauchy
estimators.

\section{Conclusion and discussion}

As we have shown in Section \ref{sec:Example}, outlying curves do
occur in
longitudinal and functional datasets. When individual smoothing is feasible,
they can be handled by the robust methods alluded to in Section \ref%
{sec:FDmodels}. But when the data is sparse and irregular, individual
smoothing is unfeasible and methods that employ the raw data must be used.
One possible approach has been presented in this article. The~idea of
using $%
t$ models to derive robust estimators is not new to Statistics [see,
e.g.,~Lange, Little and Taylor (\citeyear{LaLiTa1989})], but those procedures were specifically
developed for low dimensional multivariate data. They cannot be applied
``off the shelf'' to functional or
longitudinal data, where the dimension of the covariance matrix often
exceeds the sample size. However, an adaptation of the reduced-rank approach
of James, Hastie and Sugar (\citeyear{JaHaSu2000}) provides a way to implement $t$-model estimators in
the functional data context. The~approach we have followed is the simplest
one, which is to assume that $(\mathbf{z}_{i},\bolds{\varepsilon
}_{i})$ in
(\ref{eq:raw-data-model}) is jointly $t$ distributed, and, as a
result, the $%
\mathbf{x}_{i}$s themselves have a multivariate $t$ distribution. But other
approaches are possible. For instance, it could be assumed that
$\mathbf{z}%
_{i}$ and $\bolds{\varepsilon}_{i}$ have multivariate $t$ distributions
but are independent, or even that each $\varepsilon_{ij}$ has an
independent $t$ distribution. Unfortunately, none of these assumptions imply
that the $\mathbf{x}_{i}$s have a multivariate $t$ distribution, which
complicates the theoretical study of the estimators' properties and the
derivation of the EM algorithm. Nevertheless, these alternatives are worth
further research.

\begin{supplement}
\stitle{Technical Report and Matlab code}
\slink[doi]{10.1214/09-AOAS257SUPP}
\slink[url]{http://lib.stat.cmu.edu/aoas/257/supplement.zip}
\sdatatype{.zip}
\sdescription{The pdf file contains proofs, technical
derivations and more detailed simulation results not given in the paper. The zip file contains Matlab programs implementing
the EM algotihm for Normal and $t$ reduced-rank models.}
\end{supplement}

\section*{Acknowledgments}
 The~author thanks Professor Eugen
Schoenle, who authorized the use of the child diabetes data, and Lingsong
Zhang, who provided the Internet traffic data.

\printaddresses

\end{document}